\documentstyle[prb,twocolumn,aps,psfig]{revtex}

\draft
\begin{document}
\twocolumn[\hsize\textwidth\columnwidth\hsize\csname@twocolumnfalse%
\endcsname

\title{Magnetic Properties of (VO)$_2$P$_2$O$_7$ from Frustrated Interchain 
Coupling} 

\author{G. S. Uhrig$^1$ and B. Normand$^2$}

\address{$^1$Institut f\"ur Theoretische Physik,
Universit\"at zu K\"oln, D-50937 K\"oln, Germany.}

\address{$^2$ Departement f\"ur Physik und Astronomie, Universit\"at Basel, 
Klingelbergstrasse 82, CH-4056 Basel, Switzerland. }

\date{\today}

\maketitle

\begin{abstract}

	Neutron-scattering experiments on (VO)$_2$P$_2$O$_7$ reveal both 
a gapped magnon dispersion and an unexpected, low-lying second mode. The 
proximity and intensity of these modes suggest a frustrated coupling 
between the alternating spin chains. We deduce the minimal model containing 
such a frustration, and show that it gives an excellent account of the 
magnon dispersion, static susceptibility and electron spin resonance 
absorption. We consider two-magnon states which bind due to frustration, 
and demonstrate that these may provide a consistent explanation for the 
second mode. 

\end{abstract}

]


	Vanadyl pyrophosphate ((VO)$_2$P$_2$O$_7$, abbreviated VOPO) 
\cite{rjjgj} is a low-dimensional quantum magnetic system composed of 
S = ${\textstyle \frac{1}{2}}$ V$^{4+}$ ions. These have Heisenberg 
antiferromagnetic (AF) interactions, and a singlet ground state 
\cite{rjjgj} with spin gap $\Delta = 3.1$meV. Based on susceptibility 
and neutron-scattering experiments on powders, VOPO has been compared in 
detail to spin-ladder and alternating-chain models, \cite{rbr} both of 
which were found to be consistent with the data. Rising interest in 
spin ladders, along with the assumption that the strongest exchange 
paths would be those between V$^{4+}$ ions with the smallest separations, 
led to a preference \cite{rjjgj,rbr} for the ladder conformation, and 
for several years VOPO was quoted as the first known spin ladder system. 

	This picture was conclusively debunked by Garrett and coworkers, 
who performed the first neutron-scattering measurements on aligned single 
crystals of VOPO. \cite{rgntsb} These demonstrated unequivocally that the 
strongest exchange path in the system ($J_1$) was a double V-O-P-O-V link 
through phosphate groups in the crystallographic $b$-direction, while the 
next ($J_2$) was a double V-O-V link between edge-sharing VO$_5$ square 
pyramids, also along $\hat{b}$. The V-O-V bond along $\hat{a}$ through the 
apex of 
the pyramid, believed to be the strong ladder leg, was found to be very 
weak. The strength of the V-O-P-O-V bond was verified independently 
\cite{rtngbt} in the related compound VODPO$_4 . {\textstyle 
\frac{1}{2}}$D$_2$O, where it appears in isolation. These results are 
not unexpected, because the single electron on V$^{4+}$ in a square 
pyramidal environment occupies the $d_{xy}$ orbital in the basal 
($bc$) plane of the unit. They lead to the interpretation of VOPO as a 
set of dimerized chains, with coupling ratio $\lambda = J_2/J_1 \simeq 
0.8$.\cite{rbrt}

	The same experiment obtained the most detailed measurements to date 
for a second mode with gap 5.7meV.
This was identified from a previous study \cite{rus} as a candidate 
two-magnon bound state. In addition, the interchain coupling 
was deduced to be weakly ferromagnetic (FM), a result unexpected for a 
conventional, long superexchange path. We begin from the observation 
\cite{rus,rbkj} that frustration can act to promote bound states between 
excitations, and the hypothesis that the FM interchain coupling in fact 
results from a competition between two AF couplings. We will deduce and 
justify the nature of such a frustrated, two-dimensional (2d) model for 
VOPO, and show that it provides the basis for a complete account of the 
magnetic properties of this material. 


	The structure of VOPO was determined in detail by Nguyen {\it et al.} 
\cite{rnhs} The $b$-axis alternating chains are coupled 
along $\hat{a}$ by either the V-O-V path through apical O, or by V-O-P-O-P-O-V 
pathways through two PO$_4$ groups, while $c$-axis coupling (see also Ref. 
\onlinecite{rjjgj}) proceeds through a single phosphate unit. Quantitative 
superexchange calculations, particularly for extended pathways, remain 
beyond the scope of current understanding and computer power, \cite{rtngbt} 
particularly for the V ion. We adopt here a qualitative 
approach of identifying paths coupling strongly to the V 
$d_{xy}$ orbital, and using consistency with experiment to construct 
the minimal set of necessary interactions. 

\begin{figure}[hp]
\centerline{\psfig{figure=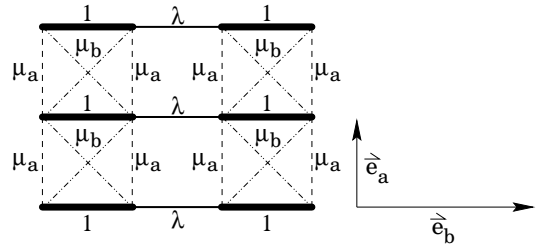,height=3.2cm,angle=0}}
\medskip
\caption{Schematic representation of the VOPO system, showing 
frustrated coupling between dimerized chains.}
\end{figure}

	We discard the possibility that the V-O-V bond along $\hat{a}$ 
is significant, because its direction is orthogonal to the 
$d_{xy}$ orbital. We argue instead in favor of the V-O-P-O-P-O-V path, 
due to the demonstrated importance of phosphate groups, and because bond 
angles throughout its length can remain close to 180$^{\rm o}$. Such paths 
connecting V ions separated only along $\hat{a}$ ($J_a$), and those 
separated both along $\hat{a}$ and by one ionic spacing along the chains 
($J_b$), are in principle rather 
similar, which presents the possibility of frustration. However, 
there are 2 types of V-O-P-O-P-O-V path, those connecting into the phosphate 
groups mediating $J_1$ and those connecting to the O atom concerned in $J_2$. 
If the latter were significant, one could expect in addition a strong 
$c$-axis coupling, which is not observed. \cite{rgntsb} 
We assume therefore that pathways of the second type are 
strongly suppressed by the angles and orbital configurations in the V-O-P 
overlap, and concentrate on the resulting 2d model in Fig. 1. 

	Here $J_1 = J$ is set to 1, $\lambda$ is defined above, and $\mu_a = 
J_a/J$ and $\mu_b = J_b/J$ parameterize the competing V-O-P-O-P-O-V 
superexchange interactions. We exclude a second-neighbor intrachain 
coupling, due to both the 
lack of a suitable exchange pathway and the minimacy criterion, as it is 
found not to be important to any of the quantities we will compute. 
That $\mu_{a,b}$ may be significant is shown by a high-temperature 
expansion \cite{ru} of the static susceptibility $\chi(T)$. This gives 
the Curie temperature $\theta_c$ as the sum of the interactions of a 
structural unit according to 
$\theta_c = - {\textstyle \frac{1}{4}} J [ 1 + \lambda + 2 \mu_a + 2 \mu_b ]$.
Comparison with the measured $\theta_c = - 84$K \cite{rjjgj} suggests that
the chain couplings \cite{rgntsb} alone satisfy at most 75\% of the sum. 

	We analyze this model using the method of Ref. \onlinecite{ru}. 
Taking the strongest ($J_1$) bonds as the sites of dimers, whose singlet 
ground states may be excited to triplets, the interactions 
permit delocalization, or hopping, of the triplets. The hopping gives a 
kinetic energy which lowers the bond singlet-triplet gap to the physical 
one, and a description of a dispersive, triply-degenerate 1-magnon excitation 
in 2d reciprocal space. While this technique is best suited to strongly 
dimerized systems, perturbation-theoretic calculations \cite{ru} show rapid 
suppression of higher-order contributions even for rather weak dimerization, 
and thus internal consistency. 

	We have calculated the coefficients for hopping of a triplet 
excitation by perturbative expansion to 5th order in the 
parameters of Fig. 1. Contributions 
beyond 3rd order indeed remain small ($\sim$3\%), and we illustrate 
the method to this order. Let $t_{ij}$ be the coefficient for the excited 
triplet to hop by $i$ dimer bonds along the chain, and across $j$ chains, 
and $\mu_{\pm}$ denote $\mu_a \pm \mu_b$, then
\begin{eqnarray}
t_{00} & = & 1 - \lambda^2 / 16 + 3\lambda^3 / 64 + 3 \mu_{-}^2 / 4
+ 3 \mu_{-}^2 \mu_{+} / 8 \label{epc} \nonumber \\ t_{10} & = & 
- \lambda / 4 - \lambda^2 / 8 + \lambda^3 / 64 + \lambda \mu_{-}^2 / 16
\nonumber \\ t_{01} & = & \mu_{-} / 2 - \mu_{-}^3 / 8 - 5 \lambda^2 \mu_{-} 
/ 32 \nonumber \\ t_{20} & = & - \lambda^2 / 16 - \lambda^3 / 64 \;\;\;\;\;\; 
t_{30} = - \lambda^3 / 128 \\ t_{02} & = & - \mu_{-}^2 / 8 - \mu_{-}^2 
\mu_{+} / 8 \;\;\;\;\;\;\;\; t_{03} = \mu_{-}^3 / 16 \nonumber \\ 
t_{11} & = & \lambda \mu_{-}^2 / 8 - \lambda^2 \mu_{-} / 32 + \lambda \mu_{-} 
\mu_{+} / 16 \nonumber \\ t_{21} & = & 3 \lambda^2 \mu_{-} / 64 \;\;\;\;\;\; 
t_{12} = - 3 \lambda \mu_{-}^2 / 32. \nonumber
\end{eqnarray}
The mode dispersion is given most straightforwardly by 
\begin{equation} 
\omega({\bf q}) = J \sum_{ij} 2^{(2-\delta_{i0}-\delta_{j0})} 
t_{ij} \cos (i q_y) \cos (j q_x), 
\label{esdr}
\end{equation}
although in fact to this order one obtains a better fit by expanding the 
(smoother) quantity $\omega^2 ({\bf q})$. 

	The measured dispersion data for the lowest-lying mode in 
Ref. \onlinecite{rgntsb}, combined with the condition on $\theta_c$, 
allow one to fit the optimal model parameters as the set 
($J, \lambda, \mu_{a}, \mu_{b}) = (10.7 {\rm meV}, 0.793, 0.203, 0.255$), 
which correspond to the superexchange interactions
\begin{equation}
(J_1, J_2, J_a, J_b) = (10.7, 8.49, 2.17, 2.73) {\rm meV} . 
\label{emis} 
\end{equation}
The various fits offer minor differences in the exact shape of the 
dispersion curve, with the higher-order calculations returning a 
very good description of the overall form. For clarity we show only 
the 5th-order fits in Figs. 2 (a) and (b) for the reciprocal-space 
cuts where data is available. The results 
(\ref{emis}) may be used to predict the form of the 1-magnon dispersion 
where it has not yet been measured, and this is shown in Figs. 2(c) and 
(d). We expect the upper band edge to be observed at 15.9meV. 
 
\begin{figure}[hp]
\centerline{\psfig{figure=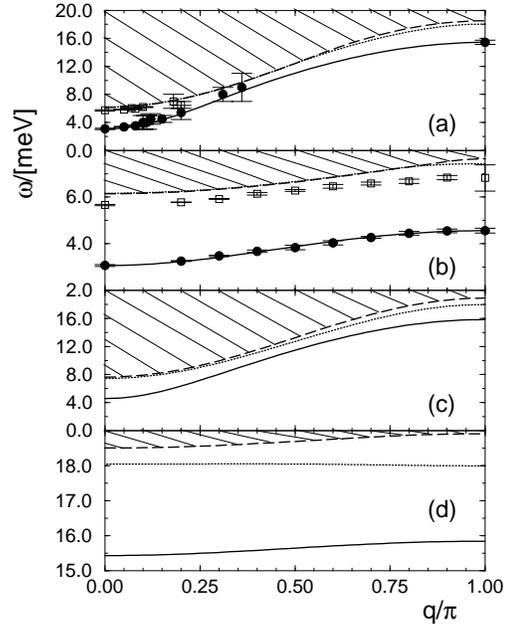,height=8.5cm,angle=270}}
\medskip
\caption{Magnon dispersion (solid lines), continuum edge (dashed 
lines) and bound $S=1$ states (dotted lines) for (a) (0,$2\pi-q$), 
(b) ($q$,$2\pi$), (c) ($\pi,2\pi-q$), and (d) ($2\pi-q,\pi$). 
Symbols are data taken from Ref. [3]. }
\end{figure}

	The primary features of the optimal parameter set are as follows. 
The alternating chains have the rather weak dimerization $\lambda \simeq 
0.8$, or $\delta = (1 - \lambda) / (1 + \lambda) \simeq 0.11$. The interchain 
couplings are significant, with a magnitude 20-25\% of the primary chain 
energy scale. The cross coupling $J_b$ is slightly larger than the pure 
$a$-axis coupling $J_a$, which is required for a net FM interchain 
dispersion, and fully plausible from above
Fig. 3 shows the density of states determined from the dispersion. The 
dominant feature is the logarithmic singularity at the 
saddle point $(\pi, 0)$, with energy 4.55meV. This energy, and not 
simply the minimum gap $\Delta$ = 3.1meV at (0,0), will play a 
significant role in determining thermodynamic quantities. 


	We next consider two further magnetic properties for 
which independent measurements are available, to determine the consistency 
of the model and parameters. The static, uniform susceptibility $\chi(T)$ 
was first measured for powder samples, \cite{rjjgj}
and more recently for single crystals. \cite{rpbaswll} The susceptibility 
to applied field $H$ measures available excitations with $\Delta S$ = 1, 
so is primarily a probe of the 1-magnon branch. It is given in general 
by $\chi = - \partial^2 F / \partial H^2 $, where $F = - \beta^{-1} \ln Z$ 
is the free energy and $\beta^{-1} = k_B T$. In the dimer model, an excited 
triplet is 
effectively a hard-core boson, because only one such state is physically 
possible on each bond. Thus thermodynamic calculations 
require a correction from Bose statistics to exclusion statistics 
appropriate for triply degenerate excitations. The free energy per dimer 
is \cite{rttw} 
\begin{equation}
f = - \beta^{-1} \ln \{ 1 + [ 1 + 2 \cosh (\beta h)] z (\beta)\} , 
\label{efttw}
\end{equation}
where $h$ denotes $g \mu_B H$ and $z (\beta)$ is the partition function 
for a single mode. The susceptibility per site is then 
\begin{equation}
\chi_0 = \beta z (\beta) / \left( 1 + 3 z (\beta) \right), 
\label{esttw}
\end{equation}
in which the denominator suppresses the Bose result at high $T$ by 
excluding multiple mode occupation.

\begin{figure}[hp]
\centerline{\psfig{figure=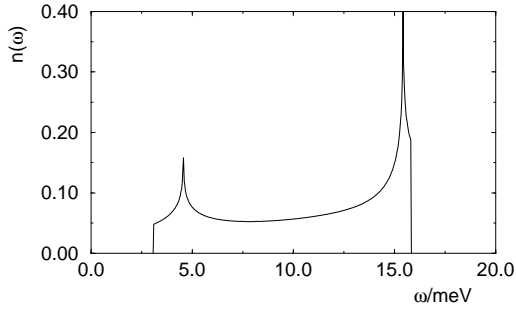,height=3.8cm,angle=270}}
\medskip
\caption{Density of states for computed magnon dispersion.}
\end{figure}

	While the statistical factor corrects for state availability, it 
does not take account of magnon interaction effects. These we may include 
within the following mean-field scheme. 
Approximating any interaction term with a spin on a neighboring 
dimer by $J_{i,i+\delta} {\bf S}_i {\bf \cdot S}_{i+\delta} \rightarrow 
J_{\delta} \langle S_{i}^z \rangle S_{i + \delta}^z \equiv J_{\delta} m 
S_{\delta}^z$, the instantaneous magnetization per site 
$m$ contributes to an effective internal magnetic field 
\begin{equation}
h_{int} = - m J \lambda - 2 m J (\mu_a + \mu_b) \equiv - m C. 
\label{eeimf}
\end{equation}
The susceptibility is defined by $m = \chi h_{ext}$, while the magnetization 
$m = \chi_0 h_{tot}$, where $h_{tot} = h_{ext} + h_{int}$ is the total field 
at each site. Simple rearrangement yields 
\begin{equation}
\chi = \chi_0 / (1 + C \chi_0), \;\;\;\; C = J ( \lambda + 2 \mu_{+} ),
\label{esmfttw}
\end{equation}
as the mean-field, interaction-corrected susceptibility.

	In Fig. 4(a) is shown the quantity $\chi(T)$ (\ref{esmfttw}).
The agreement with both powder and single-crystal data is good. Qualitatively, 
both the exclusion statistics factor and (Fig. 4(b)) the mean-field 
correction are required to deduce $\chi(T)$ within this framework. 
Quantitatively, the temperature $T_{max}$ of the peak in the 
model is 58K, rather lower than both data sets, while $\chi_{max}$ 
is fractionally smaller. This latter result corroborates
the presence of frustrating interactions, whose mean-field correction 
gives a suppression. From Fig. 4(b), the model $\theta_c$ is in excellent 
accord with the data. 

\begin{figure}[hp]
\centerline{\psfig{figure=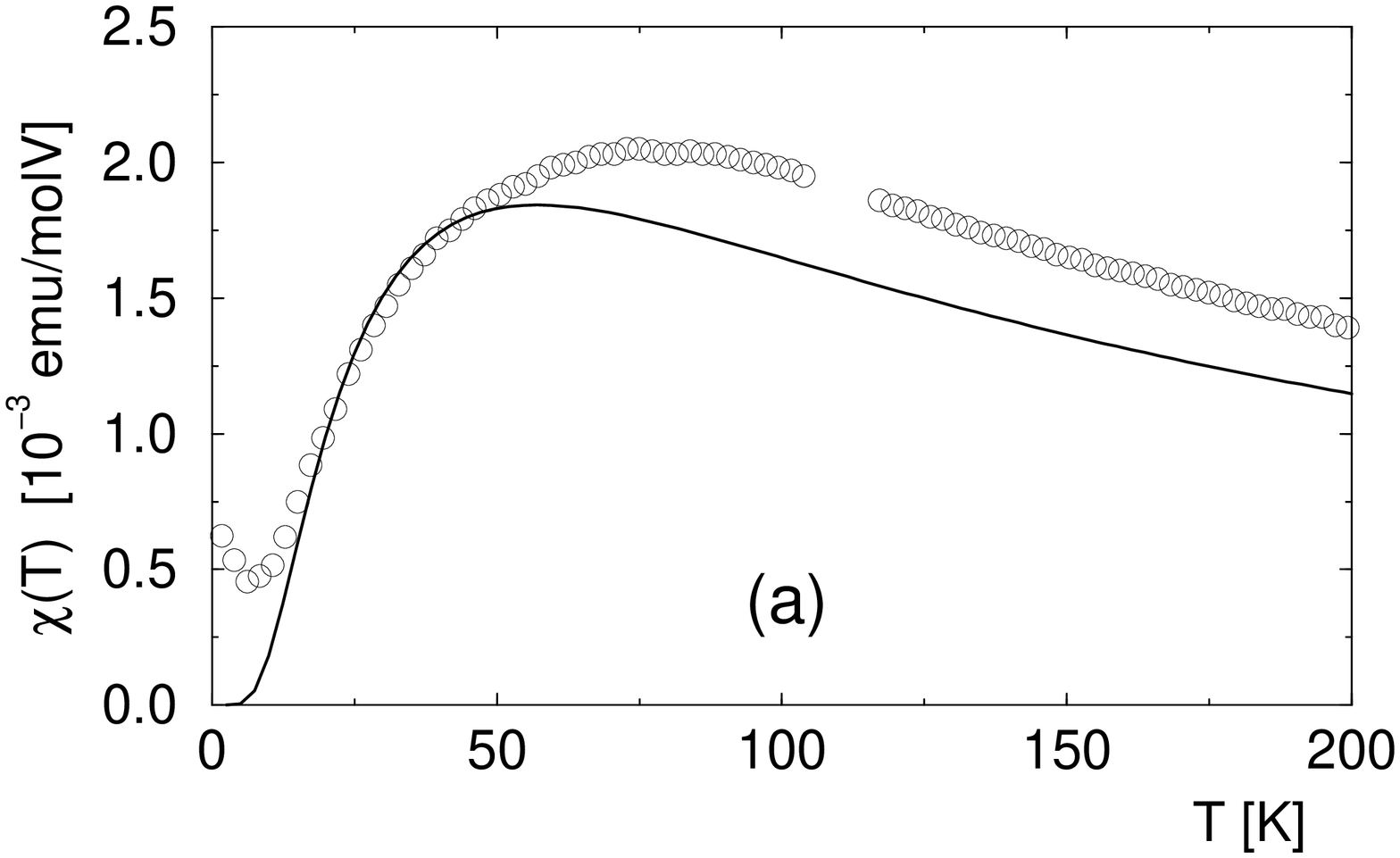,height=4.0cm,angle=270}}
\centerline{\psfig{figure=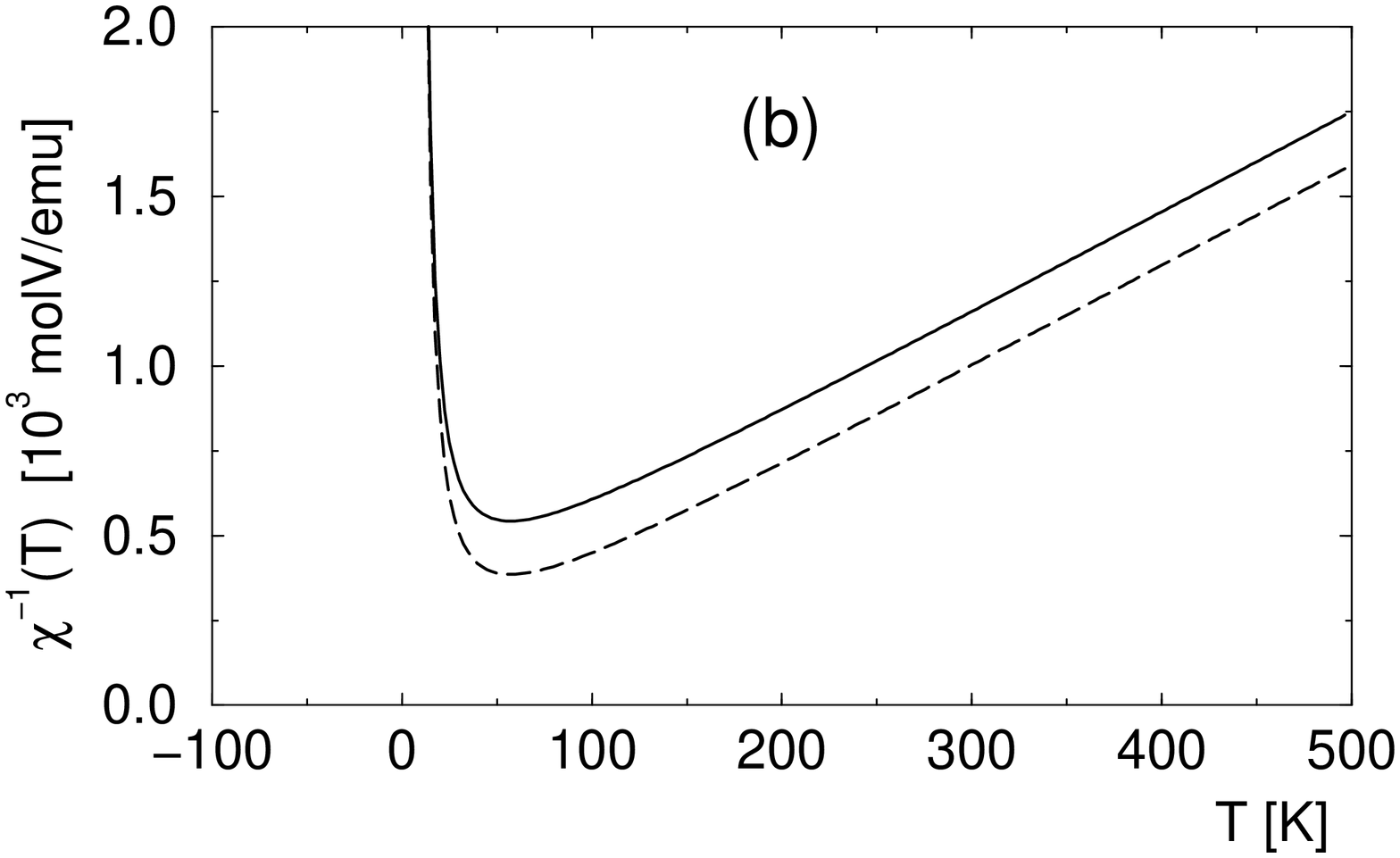,height=4.0cm,angle=270}}
\medskip
\caption{(a) $\chi(T)$ from model parameter set. Deviations from 
measured data\cite{rpbaswll} are discussed in text. (b) $\chi^{-1} (T)$ with 
(solid line) and without (dashed) mean-field correction. }
\end{figure}

	The discrepancies between calculation and experiment may be 
analyzed by applying the triplet hopping theory with mean-field 
correction to the dimerized chain. For a chain with $\lambda = 0.8$, the 
resulting $\chi(T)$ is directly comparable with essentially exact numerical 
simulations. \cite{rbbbj,rbr} 
This comparison (with Figs. 2 of Refs. \onlinecite{rbbbj} and 
\onlinecite{rbr}) reveals that $T_{max}^m / T_{max}^n \simeq 0.75$, while 
$\chi_{max}^m / \chi_{max}^n \simeq 1.2$, where superscripts $m$ and 
$n$ denote model and numerical results respectively. The first ratio is 
very similar to the discrepancy between model and experiment above. The 
second suggests further that in 1d the mean-field correction is not 
sufficient to reproduce interaction-induced suppression of $\chi$, whereas 
improved agreement is achieved in 2d. We conclude that deviations from the 
data arise primarily because the calculational scheme cannot fully account 
for magnon interactions, and not due to any intrinsic shortcomings of the 
model.


	ESR experiments have also been performed on powder and 
crystalline VOPO samples, and we focus on the latter. \cite{rpbaswll} 
For linearly-polarized (microwave) radiation, which 
excites transitions of $\Delta S^z = \pm 1$, the power absorption is 
given \cite{ra} by the imaginary part of the susceptibility 
\begin{equation} 
\chi^{\prime\prime} (\omega) = \pi \delta (\hbar \omega - h) \sinh (\beta 
\hbar \omega ) \int \frac{d^2 k}{(2 \pi)^2} e^{- \beta \omega ({\bf k})}.
\label{eesrc}
\end{equation}
Including the hard-core boson constraint as above, the intensity at the 
resonance frequency $\hbar \omega = h$ is 
\begin{equation} 
I(\beta) \propto \frac{\sinh (\beta h) z(\beta)}{ 1 + (1 + 2 \cosh (\beta 
h)) z(\beta)} . 
\label{eesri}
\end{equation}
This quantity contains in principle the mean-field correction in the form 
$h = h_{ext} - C m$. However, for the strongest resonance at 134GHz $\sim$ 
7K, the sinh function is linear at temperatures on the order of $\Delta$, 
and alterations of $h$ affect only the prefactor. 

\begin{figure}[hp]
\centerline{\psfig{figure=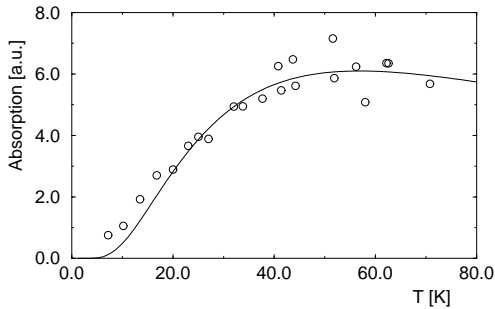,height=4.2cm,angle=270}}
\medskip
\caption{ESR absorption from model parameter set, showing excellent 
consistency with observation.\cite{rpbaswll} }
\end{figure}

	The calculated ESR absorption at 134GHz is shown in Fig. 5 with 
the data of Ref. \onlinecite{rpbaswll}. The excellent agreement indicates 
the importance of the 2d density of states, in particular the energy 
of the logarithmic singularity (Fig. 3). While the ESR absorption is often 
considered to measure a $q = 0$ property akin to $\chi(T)$, we note here a 
significant difference between the two in experiment. Theoretically, the 
triplet hopping approach yields a single energy scale for both quantities, 
which is in more satisfactory agreement with the ESR absorption, 
indicating that this is rather less sensitive than $\chi(T)$ to magnon 
interactions.


	We conclude with a brief analysis of possible bound states of 
two magnons with the computed dispersion, to seek an explanation for the 
observed second mode. The initial perturbative problem, represented 
generally as $H = H_0 + \alpha H^{\prime}$, may be brought by a continuous, 
unitary transformation \cite{rw} to a new Hamiltonian $H_{\rm eff}$ which 
conserves triplet number ($[H_{\rm eff},H_0] = 0$). This procedure is 
conducted perturbatively in $\alpha$ to generate a series for $H_{\rm eff}$.
The action of $H_{\rm eff}$ on the 
sector with one triplet gives the one-magnon dispersion, while on the 
two-triplet sector with total spin S = 1 it yields the two-magnon continuum 
and contains in addition magnon interactions. At fixed momentum one obtains 
a 2-body problem soluble by L\'anczos tridiagonalization, in particular for 
the energy of (triplet) bound states. The number of interaction 
coefficients rises very rapidly with order $n$ in $\alpha$. We have 
completed calculations for $n = 3$ and 4, finding only small alterations
($\approx 2\%$ of the continuum edge energy), but caution that these remain 
potentially significant.

	The predicted dispersion curves for the two-magnon bound state, 
computed at 4th order, are shown as the dotted lines in Fig. 2. We find 
that with the parameters fixed as above, the bound state 
does appear as an excitation mode below the continuum over a large part
of the Brillouin zone. However, this is not the case close to the band 
minimum, where it is lost in the continuum. While we thus do not always 
find a second mode below 2$\Delta$ in a consistent treatment, the discrepancy 
between the computed bound state and the observed resonance is not large:
it is reasonable to postulate that forthcoming refinement of this
calculation, and possible extension of the model to include further 
interactions, will indeed yield a quantitative explanation of the second 
mode. 


	In summary, we have presented a model for frustrated, 2d coupling 
in (VO)$_2$P$_2$O$_7$. The model gives an excellent and fully consistent 
account of the available data concerning elementary excitations, namely 
the 1-magnon dispersion, static susceptibility and ESR absorption. We use it 
to predict the dispersion in the entire Brillouin zone, and to indicate the 
origin of the observed, low-lying second excitation mode as a triplet 2-magnon 
bound state. 

	During completion of this work, we became aware of the 
contribution of Weisse {\it et al.} \cite{rwbf} These authors 
investigated the same model, albeit without the physical justification 
presented here, by diagonalization on small (up to 4$\times$8) clusters. 
Their results are qualitatively in good agreement with our expectation that 
frustration should promote bound states. Quantitatively, the parameters 
chosen differ considerably, and we would not expect to see good agreement 
with susceptibility and ESR data.  


We are grateful to U. L\"ow, B. L\"uthi, S. Nagler, H. Schwenk, and 
particularly D. A. Tennant for invaluable discussions and provision 
of data. We thank also E. M\"uller-Hartmann for helpful 
conversations, and C. Knetter for assistance with the calculation of 
$H_{\rm eff}$. GSU was supported by an individual grant and by SFB 341 
of the DFG. BN wishes to acknowledge the generosity of the Treubelfonds.

\end{document}